# Observation of Mesoscopic Clathrate Structures in Ethanol-Water Mixtures

Wei-Hao Hsu, Tzu-Chieh Yen, Chien-Chun Chen, Chih-Wen Yang, Chung-Kai Fang, and Ing-Shouh Hwang*

Institute of Physics, Academia Sinica, Nankang, Taipei 11529, Taiwan

**Abstract**

Water-alcohol mixtures exhibit many abnormal physicochemical properties, the origins of which remain controversial. Here we use transmission electron microscopy (TEM), nanoparticle tracking analysis (NTA), and atomic force microscopy (AFM) to study ethanol-water mixtures. TEM reveals mesoscopic clathrate structures with water molecules forming a crystalline matrix hosting a high density of tiny cells. The presence of these mesoscopic clathrate structures is further supported by a refractive index of 1.27±0.02 at 405 nm measured via NTA and the hydrophilic nature of the mesoscopic structures implied by AFM observations, explaining many long-standing puzzles related to water-alcohol mixtures.

Alcoholic beverages such as beer and wine have been popular for thousands of years; the major components of these beverages are water and ethanol. Both water and alcohol (monohydric alcohols such as methanol and ethanol) are liquid and highly miscible with each other under ambient conditions. Interestingly, many abnormal physicochemical properties of the mixtures (relative to ideal solutions) have puzzled researchers for decades [1–8]; anomalies are prominent in the low-alcohol concentration regime in many cases. If two liquids are miscible and form a homogeneous mixture, then in principle the total entropy of the mixture should be higher than that of the pure components prior to mixing; in reality, water-alcohol mixtures exhibit large negative excess entropy [1]. It is well known that pure liquid water forms a fluctuating three-dimensional (3D) network of hydrogen bonds, suggesting that mixing a small percentage of alcohol molecules into water would lead to endothermic destruction of water-water hydrogen bonds. Surprisingly, mixing water with alcohol leads to excessive heat evolution [1]. In 1945, Frank and Evans interpreted the observed negative excess entropies and enthalpies of water-alcohol mixtures in terms of an 'iceberg' structure that forms in the water surrounding alcohol hydrophobic groups in solution [3]; this concept has prompted many researchers to propose the formation of various molecular associates including clathrate-like structures [2,4]. However, other experiments have not supported such a scenario, with some researchers proposing incomplete mixing or segregation of water and alcohol at the molecular level [5–8]. These early studies focused on interactions among water and ethanol molecules in order to explain the observed physicochemical abnormalities.

Another puzzle is the mesoscale objects of length scale ~100 nm in water-alcohol mixtures at low alcohol concentrations, as revealed by static light scattering, dynamic light scattering, and nanoparticle tracking analysis (NTA) [9–16]. These mesoscopic structures—which are considerably larger than the molecular sizes of water and alcohol—have prompted vigorous debates as to their origin. Sedlák et al. attributed them to the association of alcohol molecules with water into supramolecular structures or complexes [9]. More recently, many groups have pointed to bulk nanobubbles, nanometer-scale gas bubbles suspended in aqueous solutions, because numerous experiments have shown that a sufficient amount of dissolved gas in solution is essential for the presence of the mesoscopic structures [10,14,15]. This possibility is further supported by two observations. First, air solubility in alcohols is considerably higher than that in water. Second, the mixing of alcohol and water causes supersaturation, as air gas is more soluble in the individual components than in the mixture [17]. However, some experimental evidence contradicts the formation of mesoscopic-scale gas bubbles: some researchers attribute the structures to trace water-insoluble impurities or organic contaminants in the mixtures [11–13]. Explanations based on bulk nanobubbles also suffer from other drawbacks. First, there is still no satisfactory theory of the observed long-term stability of bulk nanobubbles (stable for hours or even months) [16,18–20]. Second, formation of gas bubbles is not consistent with large negative excess entropy and negative enthalpy upon mixing water with alcohol.

To resolve the above mysteries, here we utilized transmission electron microscopy (TEM), NTA, and atomic force microscopy (AFM) to study ethanol-water (EW) mixtures at ~10% volume fraction ethanol (~3.3% mole fraction ethanol). TEM of EW mixtures sandwiched in graphene liquid cells revealed mesoscopic clathrate hydrate structures resembling the clathrate structures observed in gas-supersaturated water at room temperature [21]. NTA indicated a high concentration of nanoparticles 50-120 nm in size in 10% EW mixtures; these nanoparticles scattered laser light weaker than silica particles of similar sizes, indicating that the refractive index (RI) of the mesoscopic structures is close to that of water. AFM demonstrated that the mesoscopic structures were not adsorbed on highly ordered pyrolytic graphite (HOPG), a hydrophobic substrate, underscoring the hydrophilic nature of the mesoscopic structures. The mesoscopic clathrate structures uncovered with TEM represent the low-energy and low-entropy structures that explain the negative enthalpy and appreciable entropy loss detected previously, as well as many other experimental observations, which occur upon mixing water with alcohol.

We encapsulated 10% EW mixtures between two laminated graphene layers and studied the mixture with bright-field and dark-field imaging at room temperature (see Materials and Methods in Supplemental Material [22]). Encapsulated regions appeared with darker contrast than non-encapsulated regions; graphene wrinkles (dark lines in the image) often appeared in encapsulated regions (Fig. S1a in Supplemental Material [22]). In most of the encapsulated regions, mesoscopic structures featuring a high density of tiny cells with a cell separation of 5-7 nm were evident, but occasionally there were detected cells of larger sizes or irregular shape (Fig. S1b-d). Selected area electron diffraction (SAED) patterns acquired on the mesoscopic structures exhibited diffraction spots other than those corresponding to the graphene's lattices and wrinkles (Fig. S1e-g), indicating the presence of crystalline structures in the mesoscopic structures. Diffraction spots with interplanar spacing (d-spacing) of 2.6-2.9, 3.6-4.0, or 4.5-4.8 Å were commonly observed in the mesoscopic structures; these values are similar to those reported for clathrate structures in gas-supersaturated water [21].

Fig. S2a (see Supplemental Material [22]) shows TEM of another mesoscopic structure in a 10% EW mixture with the SAED shown in Fig. S2b; the cells appeared as white spots at underfocus and as dark spots at overfocus, with low contrast at focus (Fig. S2c-e in Supplemental Material [22]). We performed dark-field TEM to reveal the regions that contribute to certain diffraction spots; the sample was tilted to certain angles such that the diffraction spots related to the mesoscopic structure exhibited strong intensity (Fig. 1). Strong diffraction spots with a ring around each of them were evident (Fig. 1b-c). The dark-field image acquired from the diffracted beam (d-spacing of 2.8 Å; labeled "1" in Fig. 1b) revealed honeycomb-like structures with bright features surrounding the tiny cells (Fig. 1d); we observed many such honeycomb-like structures in dark-field images of mesoscopic structures in 10% EW mixtures. Notably, the honeycomb-like structures possessed neither long-ranged translational nor orientational order. The ring around a diffraction spot is likely caused by the honeycomb-like superlattices of the crystalline structure associated with the diffraction spot. The radius of each ring was ~0.19 nm$^{-1}$ (Fig. 1b-c), corresponding to a d-spacing of ~5.3 nm of a hexagonal superlattice or 6.1 nm of the average cell-cell spacing. Electron energy loss spectroscopy of a mesoscopic structure returned spectra of a strong carbon K-edge associated with graphene and an oxygen K-edge associated with water (Fig. 1e). Oxygen spectra were not detected in regions with no encapsulated materials (data not shown). We therefore conclude that the mesoscopic structures may consist of a type of clathrate hydrates with water molecules forming crystalline structures surrounding the tiny cells.

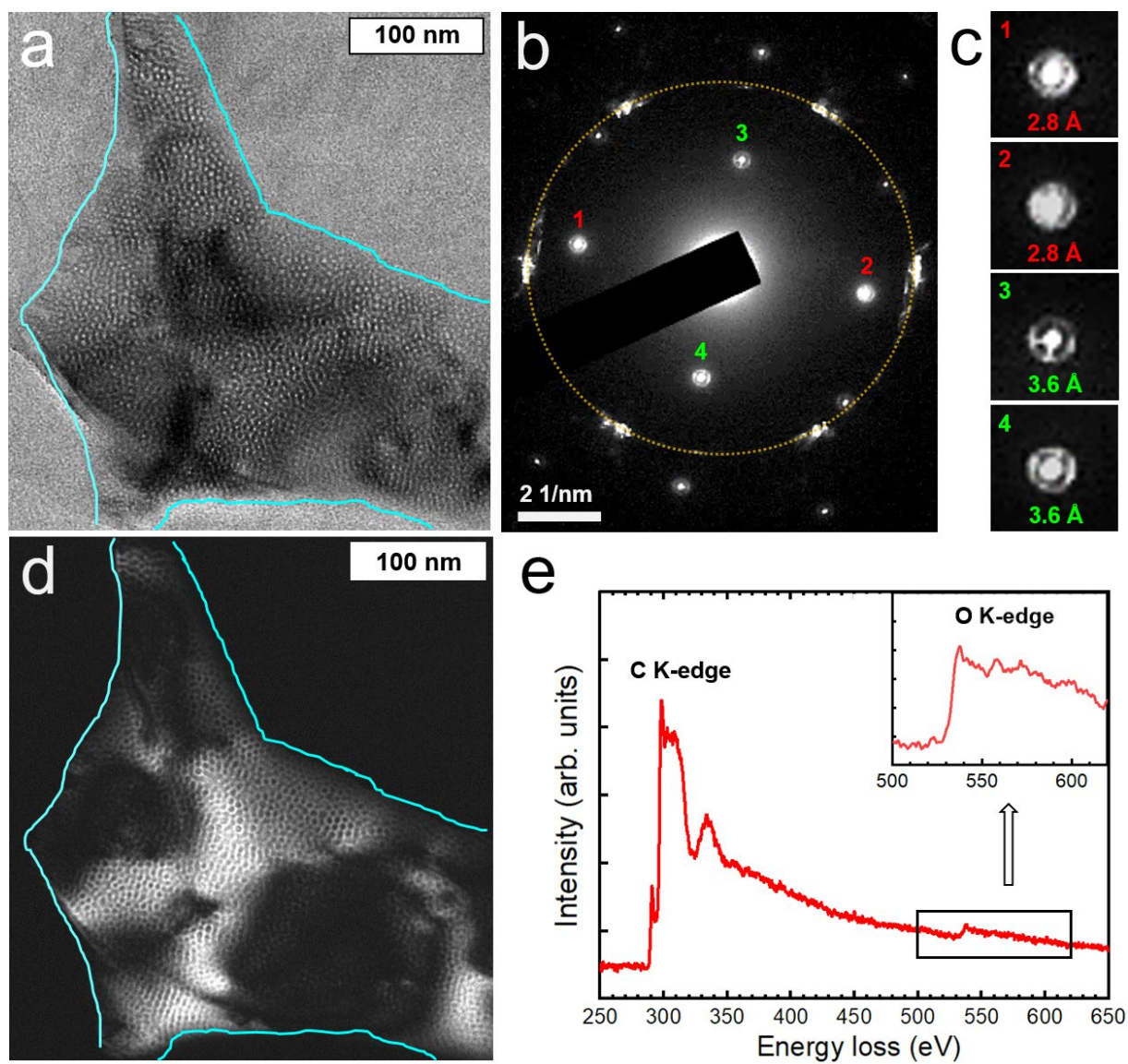

FIG. 1. TEM characterization of a mesoscopic structure in a 10% EW mixture. The region outlined in white in Fig. S2a was imaged after the sample was tilted by +14.9° around the x-axis and by -6.5° around the y-axis. (a) Bright-field TEM image. The light-blue line outlines the pocket containing the mesoscopic structure. (b) Electron diffraction recorded on the region shown in (a). Orange circle, first-order diffraction spots of the multilayer graphene. (c) Contrast-adjusted diffraction spots as numbered in (b). The d-spacings of some of the diffraction spots associated with the mesoscopic structures are indicated. (d) Dark-field TEM image acquired from the diffraction spot labeled "1" in (b). Note that the crystals that contribute to the diffracted beam appear bright. (e) Electron energy loss spectroscopy of the region shown in (a). Inset: Background-subtracted electron energy loss spectroscopy showing an oxygen K-edge.

The clathrate structures in these EW mixtures strongly resemble those in gas-supersaturated water [21] in various ways, such as their appearance in bright-field and dark-field images as well as the d-spacing of the clathrate structures. In addition, the detailed cell configuration of the clathrate structures changed slowly over time, and the change rate increased with increasing electron dose rate. We therefore imaged the structures at magnifications lower than ×40,000 (dose rate ~8.1 e$^-$Å$^{-2}$ s$^{-1}$) to minimize the effects of electron irradiation. At this low dose rate, the clathrate structures usually withstood the electron irradiation for the duration of TEM. Interestingly, cell sizes and separation became more uniform after continuous electron illumination (imaging) for several minutes. Fig. S3a,b (see Supplemental Material [22]) shows two dark-field images of the same region recorded at a time difference of 30 s. Even though most of the structures persisted, close examination of the structures reveals several subtle changes (Fig. S3c), indicating the constant restructuring of the hydrogen-bonded networks in the clathrate structures. We also compared the diffraction patterns recorded along a few zone axes and found large similarities between the clathrate structures in EW mixtures and those in gas-supersaturated water (details will be presented elsewhere). These studies indicate that EW mixtures and gas-supersaturated water share very similar clathrate structures and that the crystalline structures in the clathrate structures are a special type of 3D hydrogen-bonded networks of water molecules that differs from any water ice structure reported to date. Further effort will be needed to determine the atomic architecture of these clathrate structures.

Formation of clathrate structures upon mixing water with ethanol explains many long-standing puzzles and is consistent with various experimental observations. EW mixtures exhibit a sharp decrease in excess entropy and enthalpy [1] and a sharp increase in excess heat capacity [23] with increasing ethanol concentration in the low ethanol concentration regime. Vibrational spectroscopies based on Raman scattering and IR absorption techniques have also demonstrated a sharp increase in strength of hydrogen bonds with increasing ethanol concentration in the low ethanol concentration regime [2,4,14]. Formation of clathrate structures may well be the origin of these surprising observations. It is also consistent with a recent report of a density of ~0.91 g/cm$^3$ measured for the mesoscopic structures in EW mixtures and the incompressible nature of the structures upon increasing the external pressure from 1 atm to 5 atm [13]. Note that these clathrate structures are mainly composed of crystalline water molecules and should have a density close to that of ice (~0.92 g/cm$^3$).

It has been argued that gas supersaturation in EW mixtures leads to the formation of bulk nanobubbles. Thus, the mesoscopic clathrate structures detected here may be the so-called bulk "nanobubbles", with the shapes of the clathrate structures on TEM seriously distorted due to strong confinement by graphene layers. The long-term stability of mesoscopic structures in EW mixtures indicates that Oswald ripening is not at play [13–15], which favors clathrate structures over gas bubbles. A recent study reported that the removal of ethanol from EW mixtures does not affect the size distribution and the concentration of "nanobubbles" [14], indicating that ethanol molecules are not incorporated inside the crystalline structure of the clathrates. This observation also precludes the possibility of the formation of clathrate-like structures of ethanol and water proposed by some researchers [2,4,24]. In addition, the zeta potential of "nanobubbles" in the EW mixture is considerably lower than that for "nanobubbles" generated via other methods in pure water; this difference has been attributed to ethanol molecules adsorbing on the surface of the "nanobubbles" [14]. Thus, ethanol molecules may be present at the interfaces between the clathrate structures and the surrounding liquid water.

NTA captures the movement of individual scattering nanoparticles in a solution via dark-field microscopy. The size of a particle is derived by analyzing its trajectory according to the Stokes-Einstein relationship of Brownian motion; the size distribution and the concentration of nanoparticles can thus be determined. NTA can also be used to determine the RIs of nanoparticles in suspensions by measuring the size and light scattering intensity of individual particles [25]. The measured (maximum) scattering intensity from particles in focus can be described by the theoretical scattering cross section from Mie theory; it is a function of the particle size and the RIs of the scattering particles and the liquid medium if the wavelength and detection angle are known. At the same size, objects with a larger difference in RI from liquid medium would scatter light more strongly than objects with a smaller difference in RI. Hence, RI may distinguish whether the mesoscale objects in the EW mixtures are gas bubbles, the clathrate structures seen in TEM, or other impurities

When we performed >10 independent NTAs of 10% EW mixtures, most of the experiments had a high concentration of dim nanoparticles ($\sim 10^9$ particles/ml) with a size distribution of 50~120 nm (Fig. 2a). These particles almost disappeared when the 10% EW mixture was degassed in vacuum of ~0.1 atm for >8 h (Fig. 2b) or when both water and ethanol were pre-degassed before mixing (data not shown). These observations indicate that the dim particles are related to air gases dissolved in the solution, consistent with previous NTA studies of EW mixtures [14,15].

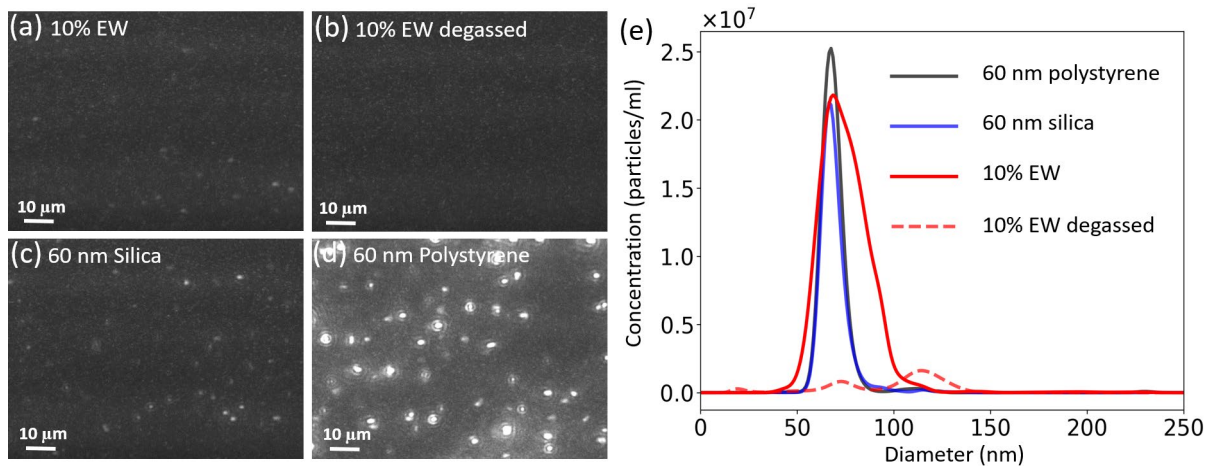

FIG. 2. NTA of a 10% EW mixture and a degassed EW mixture as well as suspensions of polystyrene and silica nanoparticles of ~60 nm in degassed EW mixtures. Images were acquired at the camera level of 16. (a-d) Typical NTA images. (e) Size distributions in the indicated experiments.

We also prepared solutions of monodisperse polystyrene and silica nanoparticles in pure water as well as in degassed 10% EW mixtures and studied these solutions with NTA for comparison with NTA of the 10% EW mixtures (Fig. 2c-d). The known sizes of these monodisperse nanoparticles and the known viscosity of pure water ($\eta_{water}$ = 0.90 cP at 25°C) [26] enabled us to calculate the viscosity of the liquid medium in the 10% EW mixture ($\eta_{EW}$) as 1.19 cP (see Fig. S4 in Supplemental Material [22]), which is consistent with that in previous studies [27,28]. Silica and polystyrene beads exhibited a very narrow size distribution around 60 nm (Fig. 2e). The mesoscopic structures in the 10% EW mixture exhibited a broader distribution with a mean diameter of ~70 nm (Fig. 2e). In general, the scattering intensity of the mesoscopic structures in the 10% EW mixture (Fig. 2a) was slightly weaker than that of silica nanoparticles (Fig. 2c), but much weaker than that of polystyrene nanoparticles (Fig. 2d). These observations suggest that the difference in RI versus liquid medium is largest for polystyrene nanoparticles and smallest for the mesoscopic structures in the 10% EW mixture.

To determine the RI of the mesoscopic structures in the EW mixtures, we adopted a method similar to that of van der Pol et al. [25] We tracked the trajectories and the scattering intensities over time of individual particles in suspensions. Due to Brownian motion, the detected scattering intensity fluctuated as a particle moved through the focal plane of the illuminating laser beam (Fig. 3a). When the particle was in focus, the maximum scattering intensity was measured and it was proportional to the scattering cross section from Mie theory with a scale factor. Note that the maximum scattering intensity was weakest for the mesoscopic structure in the 10% EW mixture and strongest for the polystyrene bead (Fig. 3a). The RIs of polystyrene nanoparticles ($n_{polystyrene}$) and pure water ($n_w$) have been measured as 1.62 and 1.34, respectively, at 405 nm (violet) [29,30]. We used these RI values and the known sizes of monodisperse polystyrene nanoparticles to determine the scale factor for Mie scattering theory; the RIs of silica nanoparticles ($n_{silica}$) and the medium in the 10% EW mixture ($n_{EW}$) were then determined to be 1.47 and 1.35, respectively (see Fig. S5 in Supplemental Material [22]). These values are consistent with previous studies [31–33], validating our method for the RI values. In Fig. 3b, the scattering intensity increased with increasing particle diameter and RI difference. Fitting with Mie theory returned RIs of mesoscopic structures of 1.27±0.02 or 1.45±0.02, with the former lower than the RI of the liquid medium (1.35) and the latter higher. The former value is favored because the mesoscopic clathrate structures seen in TEM are mainly composed of crystalline water molecules, and should therefore have a RI close to that of ice (~1.30) [34]. The smaller RI difference for the mesoscopic structure also precludes the possibility of gas bubbles ($n_{air}$~1.0), because gas bubbles would scatter laser light more strongly than would polystyrene beads of the same size (green dashed line in Fig. 3b). As the RI of most organic materials is higher than 1.48 [29], the low RI value measured here also does not support the hypothesis of mesoscopic structures of organic contaminants in EW mixtures.

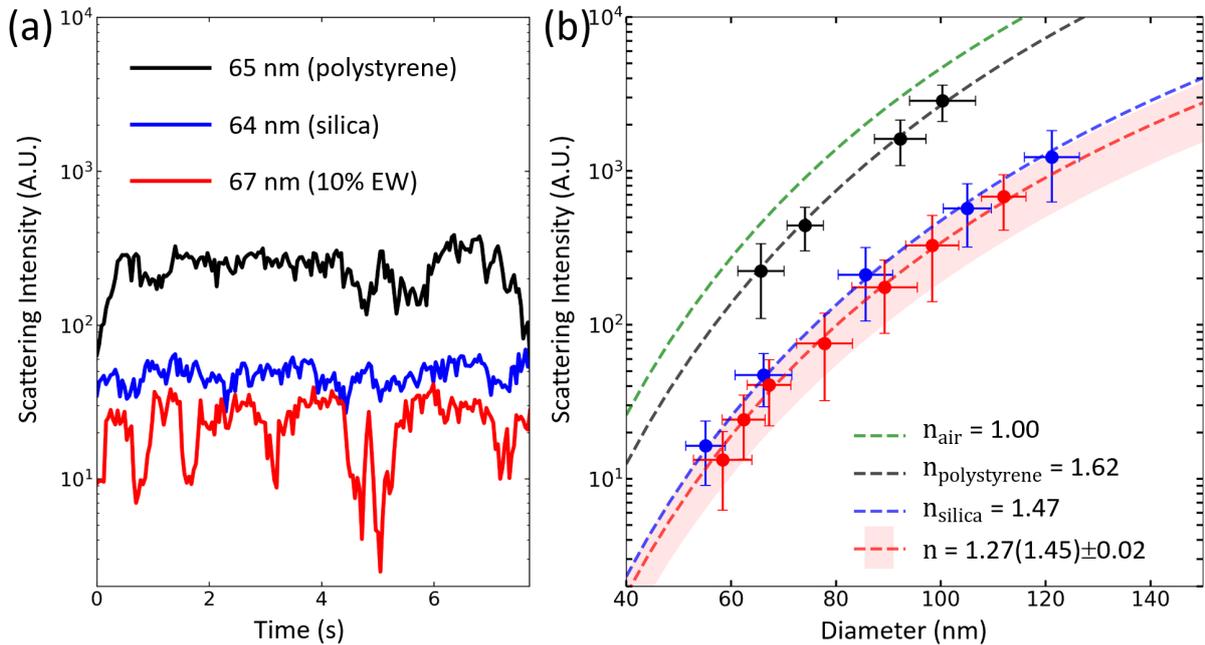

FIG. 3. Determination of the RIs of mesoscopic structures in 10% EW mixtures. (a) Scattering intensity over time of a mesoscopic structure, a silica bead, and a polystyrene bead of ~60 nm in 10% EW mixtures. The sizes of the three objects were determined by analyzing their Brownian motions. (b) Measured (symbols) and calculated (dashed lines) scattering intensity versus diameter for polystyrene, silica beads, and mesoscopic structures in 10% EW mixtures. The error bar denotes $1\sigma$ of the mean, where $\sigma$ is the standard deviation.

We deposited a 10% EW mixture on a HOPG substrate to test whether the mesoscopic structures are adsorbed onto this hydrophobic substrate; due to hydrophobic interactions, hydrophobic objects in the solutions would tend to adsorb on hydrophobic substrates. In >5 independent experiments, we did not observe any 3D cap-shaped or semispherical structure on HOPG within 3 h of deposition of the mixture (Fig. 4). This observation is consistent with a previous report of no bubbles or particles on HOPG after deposition of $CO_2$-treated EW mixtures [11]. Importantly, these AFM studies do not support the presence of hydrophobic nano-entities—such as bulk nanobubbles (with no surfactant coating), hydrophobic particles, or hydrophobic organic contaminants—in the EW mixtures, but they do support a scenario in which the nano-entities are hydrophilic clathrate structures that associate well with the surrounding aqueous solution and remain well dispersed.

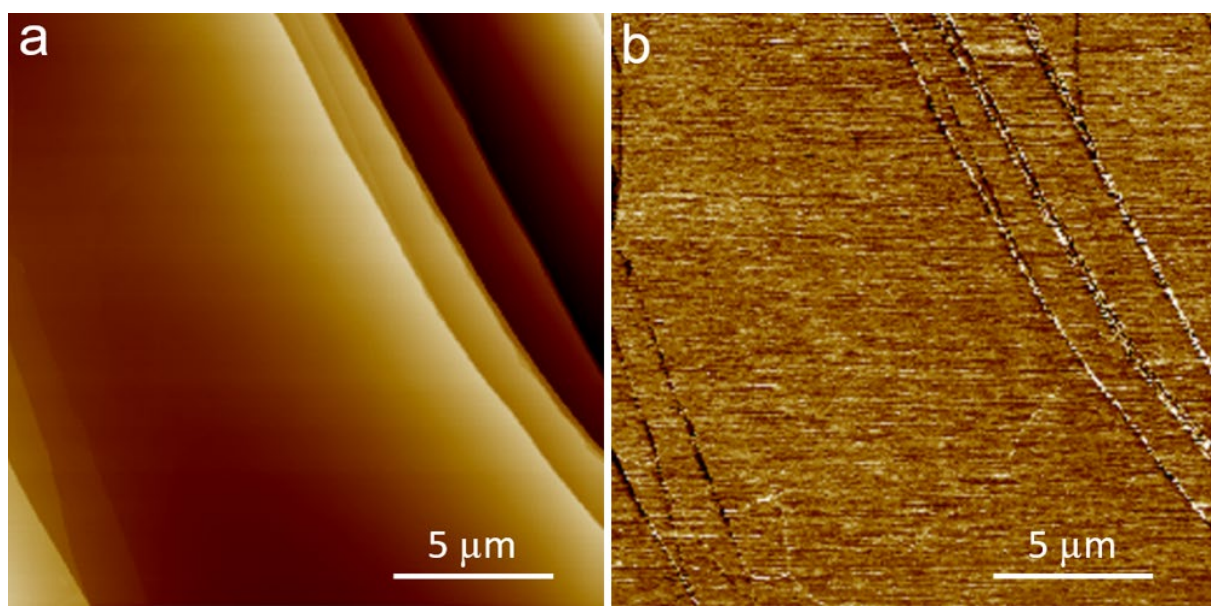

FIG. 4. AFM of an HOPG surface after deposition of an EW mixture. (a) Height image of an HOPG surface with atomic step edges. No 3D cap-shaped or semispherical structures were evident on flat terraces. (b) The corresponding stiffness map. Neither soft nor hard structures were detected on flat terraces.

Mesoscopic structures ~100 nm in size have also been detected in many aqueous solutions of small organic molecules, such as tert-butyl alcohol, tetrahydrofuran, urea, and sugar [9,10]. Jin et al. previously argued that these mesoscopic structures were bulk nanobubbles stabilized by organic molecules adsorbed at the gas-water interface [10]. Using centrifugation and light scattering, Sedlák and Rak showed that the mass densities of these mesoscopic structures in aqueous solutions of urea and tert-butyl alcohol were very close to that of water, evidencing that the mesoscopic structures are not gas bubbles [12]. Some studies also established that bulk nanobubbles do not exist in pure organic solvents [35], indicating that water is essential for forming "nanobubbles". It is thus very likely that these "nanobubbles" are clathrate structures as well. Similarly, numerous methods have been reported to generate bulk "nanobubbles" [16], but it remains controversial whether these "nanobubbles" are gas bubbles. The methodology used here, particularly TEM of solutions sandwiched in graphene liquid cells and RI measurements based on NTA, may be applied to further investigate these intriguing systems.

Early studies of water-alcohol mixtures sought interactions among water and alcohol molecules that explain abnormal physicochemical properties. The current work indicates that it may be the dissolved air gas that is responsible for many abnormalities in the mixtures. Thus it is essential to remove gas in the mixtures in order to study how water interact with alcohol molecules. Later research focused on whether nanobubbles form in water-alcohol mixtures. Here we have established that a special type of mesoscopic clathrate hydrate structures explains various long-standing mysteries. Overall, this work indicates that these mesoscopic clathrate structures may play a wide role in aqueous solutions and affect the hydrogen-bonded network of water molecules.

We acknowledge technical supports from Advanced Materials Characterization Lab in Academia Sinica, Taiwan. This research was supported by the Ministry of Science and Technology of Taiwan (MOST 106-2112-M-001-025-MY3 and MOST 109-2112-M-001-048-MY3) and Academia Sinica.

*ishwang@phys.sinica.edu.tw

# Supplemental Material for

# Observation of Mesoscopic Clathrate Structures in Ethanol-Water Mixtures

Wei-Hao Hsu, Tzu-Chieh Yen, Chien-Chun Chen, Chih-Wen Yang, Chung-Kai Fang, and Ing-Shouh Hwang*

Institute of Physics, Academia Sinica, Nankang, Taipei *11529,* Taiwan

*ishwang@phys.sinica.edu.tw

**Index of the Supplementary Material:**

**I.** **Materials and Methods**

**II.** **TEM Results**

**III.** **NTA Results**

# I. Materials and Methods

**Materials and sample preparation:** To prepare EW mixtures, ethanol (99.9%, BAKER ANALYZED, J.T. Baker) was mixed with pure water at a volume ratio of 1 to 9. We used deionized water (resistivity of 18.2 MΩ·cm) prepared using a Milli-Q system (Millipore Corp.) as pure water in TEM and commercial sterilized distilled water (NANG KUANG PHARMACEUTICAL CO. LTD.) as pure water in NTA. For preparation of degassed water, degassed ethanol, or degassed 10% EW mixtures, a tube of the liquid was placed in a desiccator, pumped to ~0.1 atm with an oil-free vacuum pump (Rocker 410, Rocker), and stored in the desiccator for >8 h. The desiccator was opened directly before use. The typical oxygen concentration for 10% EW mixtures and degassed 10% EW mixtures was ~12 mg/L and ~2 mg/L (measured by an oxygen meter, Horiba WQ-330PCD-S), respectively. In NTA experiments, we used monodisperse nanoparticles of polystyrene (Nanosphere, Thermo Fisher, MA) of diameter 60, 70, 90, and 100 nm and silica (nanoComposix, Fortis Life Sciences) of diameter 50, 60, 80, 100, and 120 nm. Monodisperse nanoparticles were diluted in pure water as well as in degassed 10% EW mixtures with a final concentration close to $10^9$ particles/ml. The preparation of graphene liquid cells has been described elsewhere [21]. Unless otherwise specified, all experiments were conducted at room temperature (22-24 °C).

**Transmission electron microscopy (TEM):** All graphene liquid cells were imaged via field-emission TEM (JEM-2100F, JEOL) operated at an acceleration voltage of 100 kV. The background pressure was ~5 ×$10^{-6}$ Pa. Unless otherwise specified, bright-field imaging was typically operated at underfocus to achieve high image contrast. Electron loss energy spectroscopy was conducted with a GIF 863 Tridiem (Gatan) in diffraction mode.

**Nanoparticle tracking analysis (NTA):** All NTA measurements were performed with a NanoSight (NS500, Malvern, software version 3.1) at 25.0 °C using a violet laser light source (70 mW, 405 nm). For each sample, 5 videos of 60 s (24.9 fps) were recorded; the videos were analyzed with NTA software (NTA Dev Build 3.2.16, Nanosight). Various camera settings (shutter and gain) were applied and calibrated to prevent pixel saturation while maintaining a sufficient number of analyzed particles.

**Atomic force microscopy (AFM):** AFM was performed on a Bruker AXS Multimode NanoScope V equipped with a commercial fluid cell tip holder. The substrate for sample adsorption was a square piece of HOPG (lateral size of 12 mm × 12 mm, ZYH; Structure Probe, Inc.). The EW mixture was injected onto a HOPG substrate in the liquid cell. The surface was then imaged with peak force tapping mode under liquid environment. Backside Au-coated Si cantilevers (Nanosensors, FM-AuD) with a resonance frequency of 22-32 kHz in solution and a spring constant of 2-4 N/m were used. Prior to AFM measurements, the AFM probe was cleaned with ultraviolet light.

## II. TEM Results

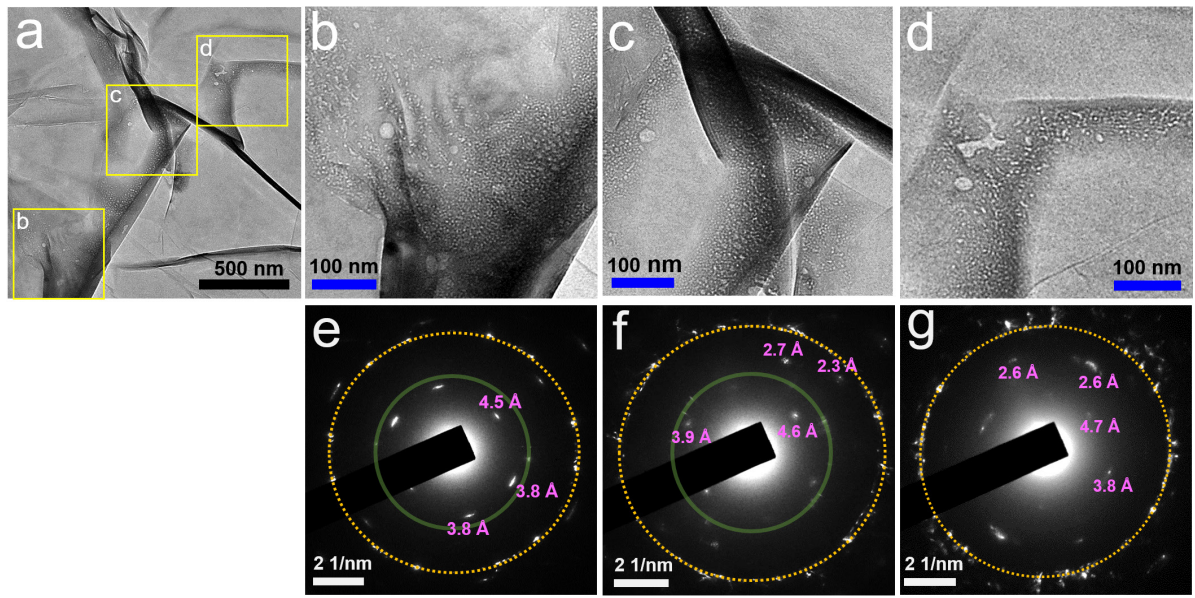

FIG. S1. TEM of an EW mixture sandwiched in a graphene liquid cell. (a) Bright-field image. (b-d) Higher-resolution images of the outlined regions labeled "b", "c", and "d" in (a), respectively. (e-g) Diffraction patterns acquired on the regions in (b-d), respectively. Orange circle, first-order diffraction spots of the multilayer graphene; green circle, diffraction spots with a d-spacing of 3.3 Å, which are associated with graphene wrinkles. Purple, interplanar spacing values (d-spacing) for some of the diffraction spots associated with the clathrate structures. Note that the diffraction spots are not commensurate with graphene's lattices.

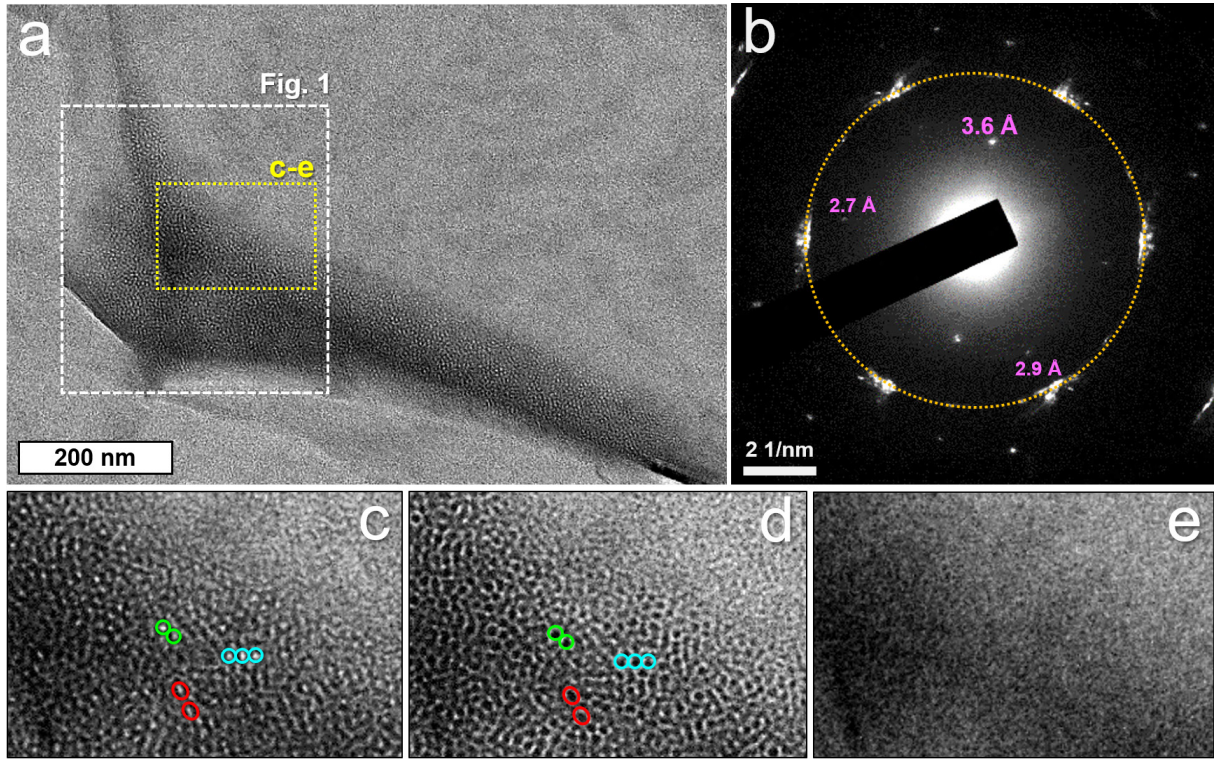

FIG. S2. TEM of a mesoscopic structure in an EW mixture. The sample was not tilted. (a) Bright-field image acquired at underfocus. (b) Selected area electron diffraction of the region in (a). Orange circle, first-order diffraction spots of the multilayer graphene; purple, d-spacing of some of the diffraction spots associated with the mesoscopic structures. (c-e) Bright-field images of the region outlined in yellow in (a) were acquired at underfocus, overfocus, and in focus, respectively. The cells appear as bright spots at underfocus and as dark spots at overfocus, as indicated with colored circles. Little contrast for the cells is seen in focus.

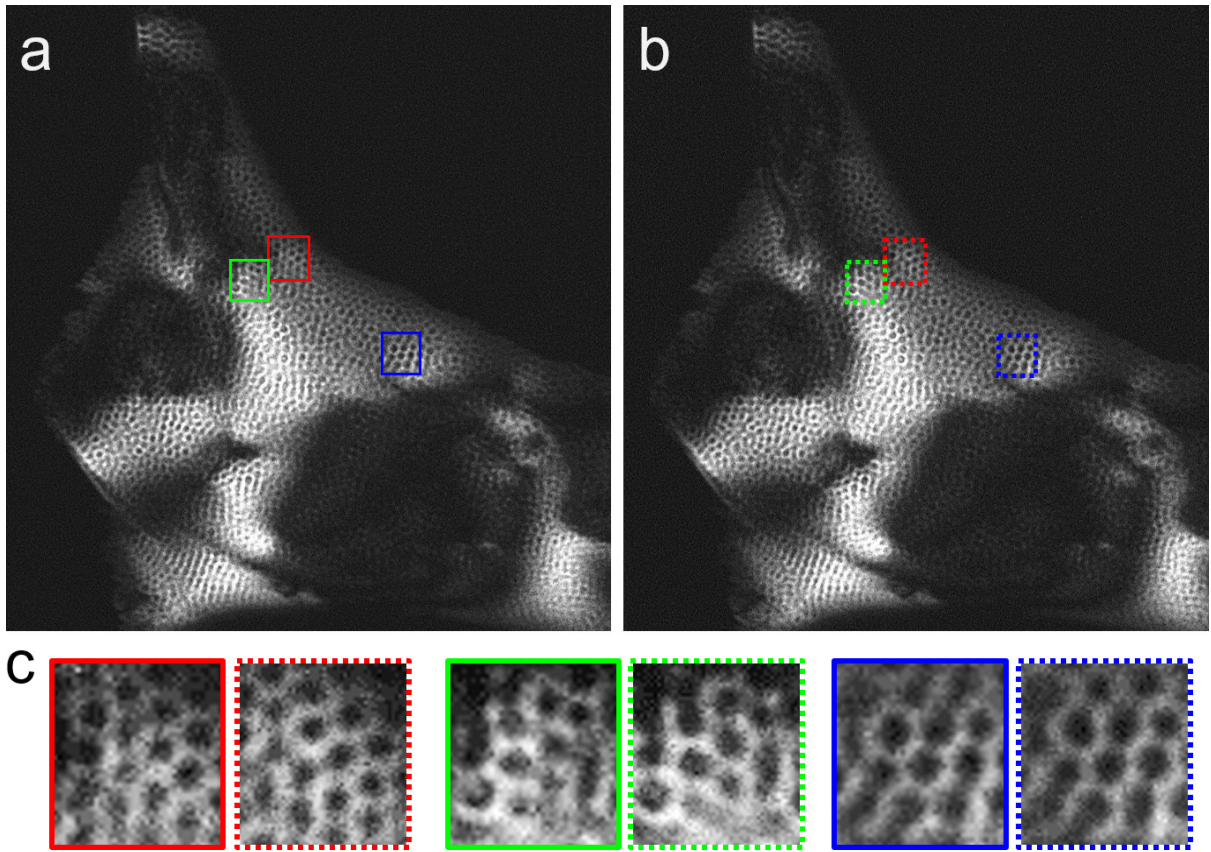

FIG. S3. Dark-field TEM images of the slow change in clathrate structures in an EW mixture. (a) Dark-field image (same as in Fig. 1c). (b) Dark-field image of the same region acquired 30 s later. (c) Comparison of zoom-in images of three outlined regions in (a) and (b).

## III. NTA Results

### 1. Determination of viscosity of 10% ethanol-water (EW) mixtures with nanoparticle tracking analysis (NTA)

Based on the Stokes–Einstein equation, we used monodisperse nanoparticles of known sizes and the known viscosity of pure water to determine the viscosity of the liquid medium in 10% EW mixtures. The Stokes–Einstein equation correlates the diffusion coefficient $D$ with the hydrodynamic radius $r$ of the particle and dynamic viscosity $\eta$:

$$D = \frac{k_B T}{6 \pi \eta r}$$

, where $k_B$ is the Boltzmann constant and $T$ is the absolute temperature. The motions of commercial monodisperse beads were recorded and analyzed through the NTA software (NTA 3.2.16), and the diffusion coefficients of individual particles were calculated based on their mean squared displacements. With a series of particles of known size, we calculated the unknown viscosity of liquid medium.

We first diluted monodisperse nanoparticles of polystyrene and silica of several diameters in pure water separately and measured the hydrodynamic radius of the polystyrene and silica beads by least-square fit based on the Stokes–Einstein equation, taking the viscosity of pure water ($\eta_{\text{water}}$) to be 0.90 cP at 25 °C. When we plotted $\frac{1}{D}$ vs. $\frac{6\pi r}{k_B T}$, all data points fell on a line with a viscosity of 0.90 cP (black dashed line in Fig. S4). The mean hydrodynamic diameters of monodisperse polystyrene beads of 60, 70, 90, and 100 nm were determined to be 65, 72, 92, and 99 nm, respectively; the mean hydrodynamic diameters of monodisperse silica beads of 50, 60, 80, 100, and 120 nm were determined to be 55, 63, 85, 106, and 123 nm, respectively. We then diluted monodisperse nanoparticles of silica in degassed 10% EW mixtures and used the above measured mean hydrodynamic diameters to determine the viscosity of 10% EW mixtures ($\eta_{\text{EW}}$). All data points fell on the best-fit red dashed line with a viscosity of 1.19 cP (Fig. S4). Five videos were recorded for analysis for each sample of monodisperse nanoparticles; the error bar denotes $1\sigma$ of the mean, where $\sigma$ is the standard deviation. The viscosity value of 1.19 cP is consistent with that in previous studies [27,28].

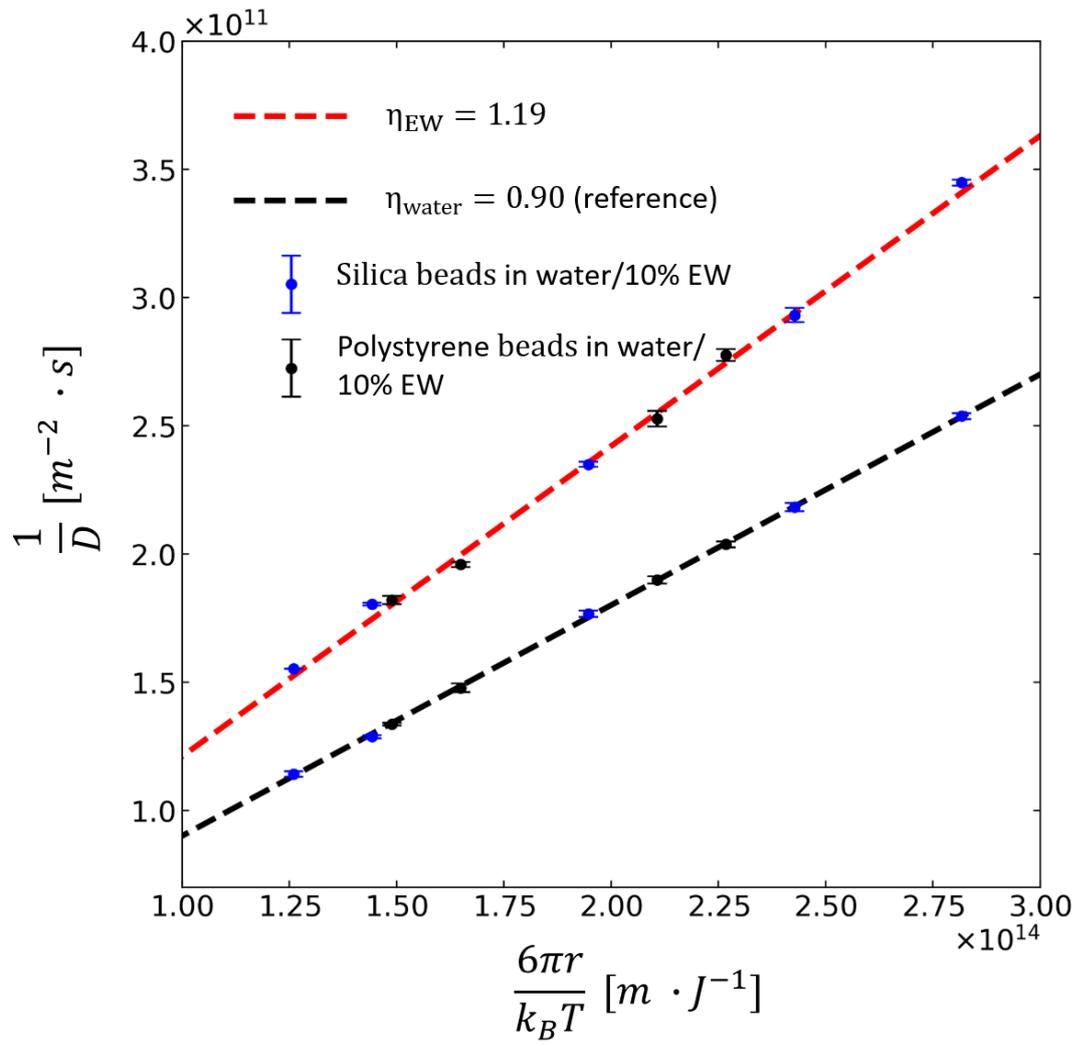

FIG. S4. Plot of $\frac{1}{D}$ versus $\frac{6\pi r}{k_B T}$.

## 2. Determination of the refractive indices (RIs) of the liquid medium and mesoscopic structures in 10% EW mixtures with NTA

The RIs of polystyrene ($n_{polystyrene}$) and pure water ($n_w$) have been measured as 1.62 and 1.34, respectively, at a wavelength of 403 nm [29,30]. We first used monodisperse nanoparticles of polystyrene (mean diameters of 65, 72, 92, and 99 nm) in pure water to determine the RIs of monodisperse silica beads (mean diameters of 55, 63, 85, 106, and 123 nm) in pure water by adopting a method similar to that of van der Pol et al.[25] The diameter and maximum light-scattering intensity of individual particles were measured by tracking their trajectories and scattering intensities over time. We developed a Python program to solve the inverse scattering problem with Mie theory and fit the calculations with the maximum scattering intensity measured from monodisperse nanoparticles of polystyrene based on the RI of 1.62 (black dashed line in Fig. S5). With the same scale factor and fitting procedure, we determined the RI of silica ($n_{silica}$) to be 1.47 at 403 nm (blue dashed line in Fig. S5), which is consistent with previous measurements [32]. We then diluted monodisperse silica beads of mean diameters of 55, 63, 85, 106, and 123 nm separately with degassed 10% EW mixtures and measured the diameter and maximum light-scattering intensity of individual silica particles. With an RI of 1.47 for silica, we determined the RI of degassed 10% EW mixtures to be 1.35 through fitting with Mie theory (light blue dashed line in Fig. S5). This value is also in agreement with previous studies [31,33], validating our method.

The RIs of mesoscopic structures in 10% EW mixtures ($n_{EW}$) were determined via the same procedure. On many occasions, the size distribution of mesoscopic structures exhibited a major peak with a certain width (Fig. 2e). We found that the mean diameter varied somewhat when a new mixture was prepared and increased slowly over time (hours) for the same mixture. For each NTA measurement, five videos were recorded. Our analysis was performed only on the mesoscopic structures with a tracklength over a minimum number of video frames and with size within one standard deviation of the mean diameter. The minimum tracklength was 30 frames for 50-60 nm structures, 50 frames for 60-70 nm structures, 80 frames for 80-100 nm structures, and 90 frames for structures larger than 100 nm. We note that an increase in minimum tracklength increases the precision of the measured diameter and scattering intensity but also reduces the number of analyzed particles. More than 100 mesoscopic structures were tracked in each data point in Fig. 3b of the main text.

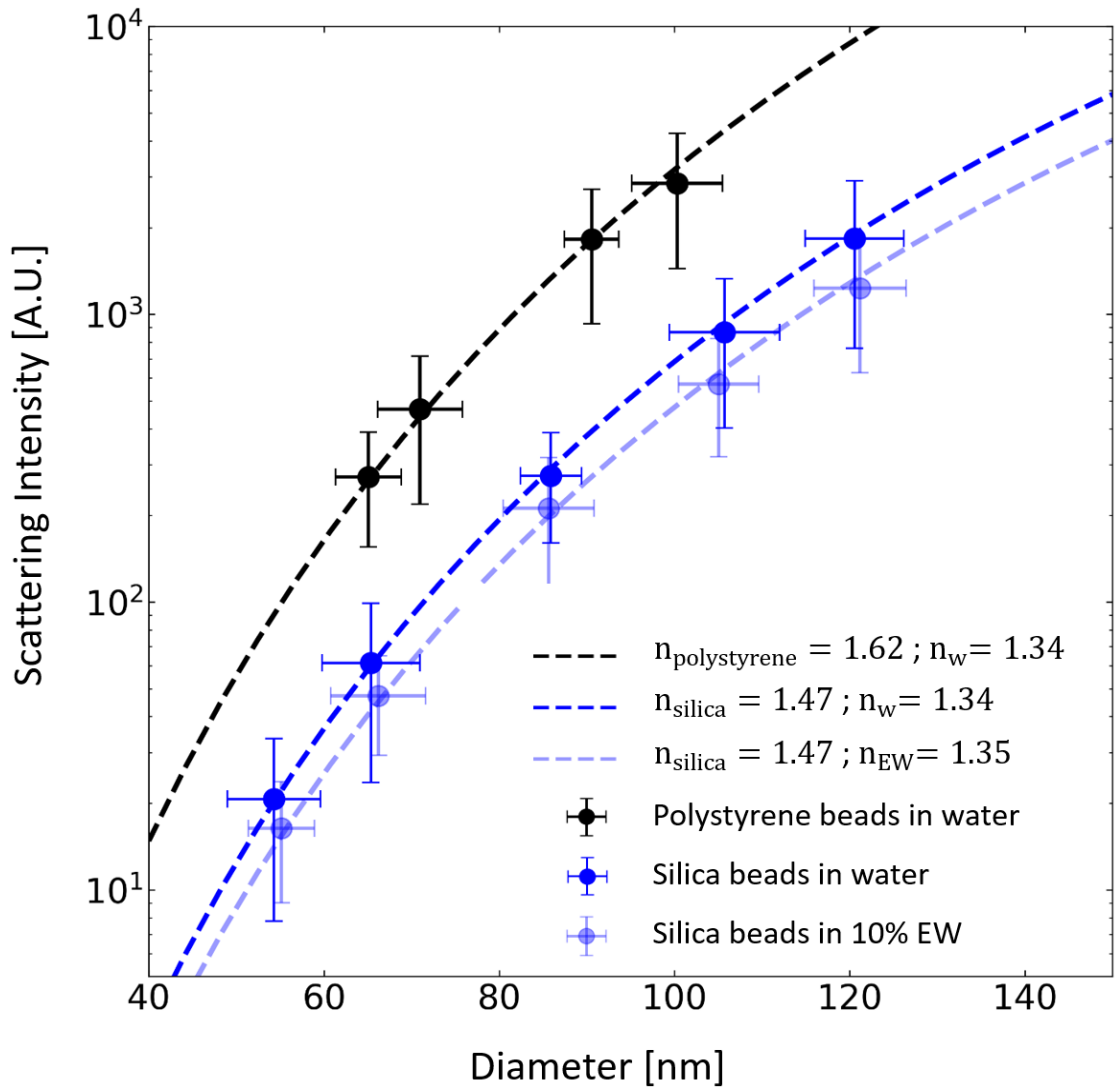

FIG. S5. Plot of measured (symbols) and calculated (dashed lines) scattering intensity versus particle diameter.